\documentclass{elsart}

\usepackage{graphics}
\usepackage{graphicx}
\usepackage{amssymb}

\begin{document}
\begin{frontmatter}

\title{High-pressure X-ray diffraction study on $\mathbf{\alpha}$-PbF$\mathbf{_2}$}

\author[label1]{L. Ehm \corauthref{cor}}\ead{ehm@min.uni-kiel.de},
\author[label1]{K. Knorr}, \author[label2]{F. M\"adler}, \author[label1]{H. Voigtl\"ander},
 \author[label3]{E. Busetto}, \author[label3]{A. Cassetta},
 \author[label3]{A. Lausi} \&  \author[label4]{B. Winkler}  

\address[label1]{Institut f\"ur Geowissenschaften,
 Mineralogie,\\  Christian-Albrechts-Universit\"at zu Kiel, \\
 Olshausenstra{\ss}e 40, D-24098 Kiel, Germany}

\address[label2]{Hahn-Meitner-Institut, Department I/DN, \\ 
Glienicker Stra{\ss}e 100, D-14109 Berlin, Germany}

\address[label3]{ELETTRA, Sincrotrone Trieste, \\
 Strada Statale 14 -  km 163,5 in AREA Science Park,\\ 
  I-34012 Basovizza, Italy}

\address[label4]{Institut f\"ur Mineralogie, Kristallographie, \\
Johann Wolfgang Goethe-Universit\"at, \\
 Senckenberganlage 30, D-60054 Frankfurt am Main, Germany}

\corauth[cor]{Corresponding author}

\begin{abstract}
  The high-pressure behaviour of $\alpha$-PbF$_2$ has been
  investigated by angular-dispersive synchrotron X-ray powder
  diffraction up to 6.55(8)~GPa.  The fit of a 3$^{rd}$-order
  Birch-Murnaghan equation-of-state gave the volume at zero pressure
  V$_0$=194.14(4)~\AA$^3$ and the bulk modulus b$_0$=47.0(6)~GPa with
  the pressure derivative b'=7.9(4). The continuous-symmetry-measure
  approach has been used for the quantification of the distortion of
  the coordination polyhedron in $\alpha$-PbF$_2$ revealing an
  increasing distortion with pressure.
\end{abstract}

\end{frontmatter}

\section{Introduction}
\label{intro}

\begin{figure}[h!]
\includegraphics[width=0.95\textwidth]{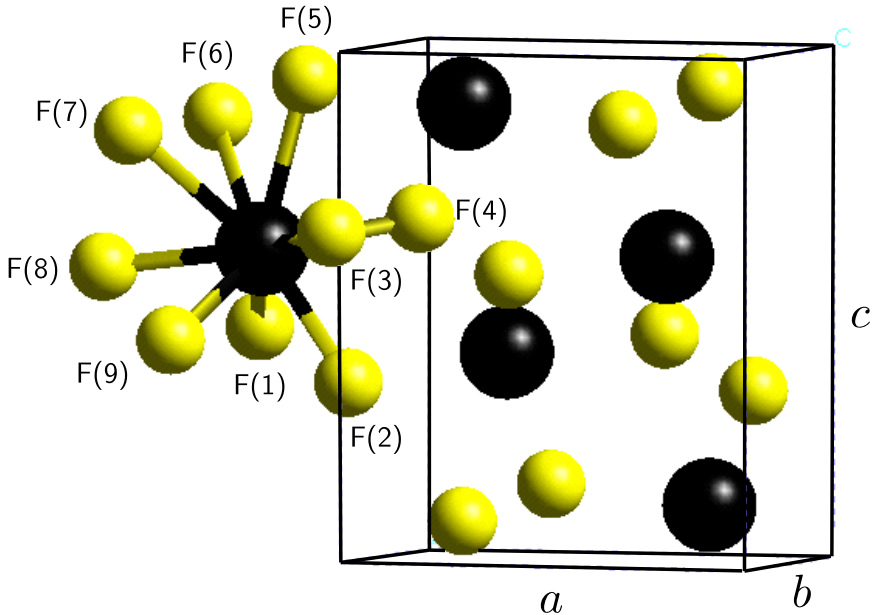}
\caption{\label{structure} Schematic drawing of the $\alpha$-PbF$_2$
  crystal structure. The black spheres represent Pb$^{2+}$ and the
  gray ones F$^-$. The first coordination shell around 
  Pb$^{2+}$ is shown where at F(1), (5), (8), (9) and F(2), (3), (4),
  (6) and (7) occupy symmetrically equivalent positions.}
\end{figure}

Lead fluoride can be found in two polymorphs at ambient conditions:
$\beta$-PbF$_2$ in the fluorite type structure \cite{Bystroem1947} and
$\alpha$-PbF$_2$ in the orthorhombic cotunnite type structure (fig.
\ref{structure}) with space group $Pnma$ and $a$=6.440~\AA,
$b$=3.899~\AA, and $c$=7.651~\AA \ \cite{Boldrini1967}. The structure
is built up by corrugated layers of tricapped trigonal PbF$_9$ prisms
along $a$ sharing an edge parallel to the $b$ direction. All atoms
occupy the Wyckoff position 4$c$ (x, 1/4, z) with x=0.2527(13),
z=0.1042(7) for Pb, x=0.8623(21), z= 0.0631(15) for F(1) and
x=0.4662(20), z=0.8457(17) for F(2), respectively \cite{Boldrini1967}.

It is not clear which polymorph is thermodynamically stable at ambient
conditions. In recent years both phases have intensively been studied
due to their potential industrial applications
\cite{Samara1976,Oberschmidt1981,Chandra1981,Aurora1990,Alov1995,Lorenzana1997,Hull1998,Costales2000}.
$\beta$-PbF$_2$ shows a high superionic conductivity with a low
superionic transition temperature suitable for solid state batteries
and sensors \cite{Chandra1981}. The $\alpha$-phase is a strong
scintillator useful in detector technology \cite{Alov1995}.  The
high-pressure behaviour of PbF$_2$ has been studied recently by
various experimental and theoretical techniques.  Lorenzana {\it et
  al} \cite{Lorenzana1997} investigated the high-pressure properties
of PbF$_2$ by Raman spectroscopy in conjunction with a first
principles, full-potential linear muffin-tin orbital total energy
method in the pressure range from ambient pressure up to 30~GPa.  The
authors detected a phase transition at approximately 0.5~GPa from the
cubic $\beta$-PbF$_2$ to the orthorhombic $\alpha$-PbF$_2$.  At
14.7(5)~GPa a second phase transition to $\gamma$-PbF$_2$ with a yet
unknown structure has been observed.  Considering previously
investigated high-pressure phases of isotype ionic AX$_2$ compounds
(e.g. SnCl$_2$, PbCl$_2$, BaCl$_2$ and BaF$_2$
\cite{Leger1995a,Leger1995,Leger1996}), Lorenzana {\it et al} proposed
a hexagonal Ni$_2$In type structure and a monoclinic structure with a
doubled $a$ axis as possible structures for $\gamma$-PbF$_2$.  Their
calculations revealed that above 16.4~GPa, PbF$_2$ in the Ni$_2$In
type structure is more stable than $\alpha$-PbF$_2$. However,
$\gamma$-PbF$_2$ in a hexagonal symmetry could not explain the
observed Raman spectra and consequently, the authors proposed the
monoclinic structure for the high-pressure $\gamma$ polymorph.
High-pressure and high-temperature neutron diffraction experiments
have been performed from ambient conditions up to pressures of 1~GPa
and temperatures of 966~K to investigate the phase diagram and to
determine the nature of the disorder of the fluorine ions at high
temperature and its pressure dependence within the $\alpha$ and
$\beta$ polymorphs of PbF$_2$ \cite{Hull1998}.  The phase stability of
PbF$_2$ under high-pressure has also been studied by {\it ab initio}
calculations using the perturbed ion method \cite{Costales2000}. The
calculations have been performed for the $\alpha$ and $\beta$
polymorphs as well as for the two proposed $\gamma$-PbF$_2$
structures. The results of the calculations support the existence of
$\gamma$-PbF$_2$ in the hexagonal Ni$_2$In type structure above
20.2~GPa.  Experimentally determined high-pressure data are mainly for
cubic $\beta$-PbF$_2$ \cite{Lorenzana1997,Hull1998}.  Hence, the goal
of the work presented here was to measure the structural changes
induced by high-pressures in order to elucidate the compression
mechanism of $\alpha$-PbF$_2$ and to extend the limited pressure range
of the study of Hull and Keen \cite{Hull1998} for a precise
determination of the equation-of-state.  Here we present the results
of a high-pressure synchrotron powder diffraction study of
$\alpha$-PbF$_2$ up to 6.55~GPa.

\section{Experimental}
\label{exp}

The X-ray diffraction experiments were performed using a commercial
(Merck, omnipur) polycrystalline sample of PbF$_2$.  By conventional
X-ray powder diffraction it was found that the sample is pure
$\alpha$-PbF$_2$.  High-pressure powder diffraction experiments up to
6.55(8)~GPa were performed at the wiggler beam-line 5.2R at the
ELETTRA synchrotron in Trieste, Italy.  The beam-line optic consists
of a Si(111) double-crystal monochromator in non-dispersive
configuration followed by a three-segment platinum-coated toroidal
focusing mirror with a horizontal acceptance of 2.8~mrad.  The
incident beam was collimated to 80~$\mu$m diameter. High-pressure
powder diffraction patterns were collected at the wavelength
$\lambda$=0.6888~\AA \ using a marresearch (mar345) image plate
detector with a pixel-size of 100~$\times$100~$\mu$m. The exposure
time per image was about 20 minutes. Pressure was applied using a
Diacell DXR-6 diamond anvil cell mounted on a $xz$-positioning-stage
for the adjustment in the incident beam.  The sample was placed in the
hole ($\varnothing$ =150~$\mu$m) of an Inconel gasket preindented to
100~$\mu$m. In order to ensure hydrostatic conditions a 16:3:1
methanol-ethanol-water mixture was used as pressure transmitting fluid
\cite{Fujishiro1982}. We used the ruby fluorescence method for the
pressure determination applying the Piermarini pressure scale
\cite{Piermarini1975}. The sample to detector distance was determined
by measuring the sample in the uncompressed cell.  Geometry parameters
for the radial integration of the two-dimensional data were determined
using {\tt FIT2D} \cite{Hammersley1996}.  For the transformation into
standard one-dimensional powder patterns the software {\tt TWO2ONE}
\cite{Vogel2002} was used, which allows for counting statistics of
multiply measured data points in order to provide proper error
estimates of the intensities and hence, a correct weighting scheme in
subsequent least-squares refinements \cite{Chall2000}.  The
one-dimensional powder patterns were corrected for attenuation by the
pressure cell.  Lattice parameters were obtained from profile matching
and the structure was refined at five different pressures by the
Rietveld method employing the program {\tt FULLPROF}
\cite{Rodriguez1993}.  The background was described by a linear
interpolation between selected points and the peak profile was
modelled using a pseudo-Voigt function \cite{Thompson1987}. The
structural parameters from Boldrini and Loopstra \cite{Boldrini1967}
were taken as the initial values for the refinements.  The isotropic
displacement parameters (U$_{iso}$) for lead and the two fluorine
atoms were determined from the measurement at ambient pressure and
assumed to be constant in the observed pressure range.  The standard
deviations of the refined parameters were scaled with the Berar-factor
\cite{Berar1991}.  The pressure-volume data were fitted by a
3$^{rd}$-order Birch-Murnaghan equation-of-state \cite{Birch1978}.
Since pressure and volume are subject to experimental errors, which
have to be weighted correctly in the fit procedure, the conventional
least-squares fit is not appropriate. Consequently, the method of
orthogonal distance regression (ODR) was used and the
equation-of-state implemented in the program code {\tt ODRPACK}
\cite{Boggs1989,Boggs1992} to minimise
\begin{equation}
\label{odr}
 \sum_{i}(w_{p,i}(p_i^o-p_i^c))^2+(w_{V,i}(V_i^o-V_i^c))^2,
\end{equation} 
with weights $w_{p,i}=c_1/\sigma_{p_i^o}^2$ and
$w_{V,i}=c_2/\sigma_{V_i^o}^2$. The parameters $c_{1,2}$ are
adjustable parameters ensuring that the weighted experimental errors
$w$ are in the same order of magnitude for both variables
\cite{Boggs1992}.  , The difference between the observed and the
calculated values for p$_i$ and V$_i$ are at least one order of
magnitude smaller than the experimental error bars, when the
experimental error bars are used as weights in the fit-procedure.
This indicates that the relative precision of the observed data
points, which influences the fit, is higher than that represented by
the error bars. Taking the differences $\Delta$p and $\Delta$V between
the observations and the equation-of-state as weighted errors in the
subsequent fitting cycle did not affect the values of the fitted
parameters themselves but their error estimates.
 
\section{Results and Discussion}
\label{res}

\begin{figure}[h!]
\includegraphics[width=1.0\textwidth]{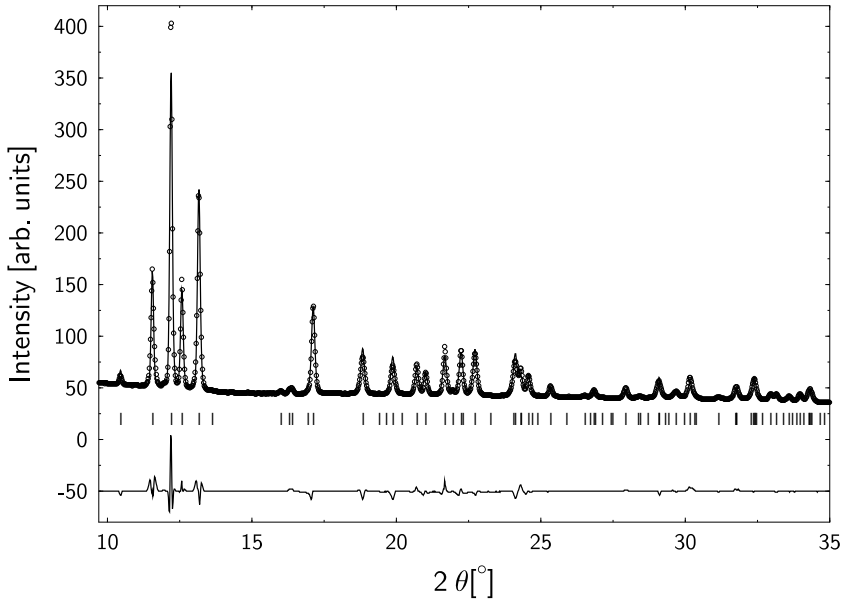}
\caption{\label{powder-fit} Observed (circles) and calculated (solid
  line) diffraction pattern for $\alpha$-PbF$_2$ at 3.8(1)~GPa.  The
  difference and the tick marks for the calculated reflection
  positions are plotted at the bottom of the figure.}
\end{figure}

\begin{figure}[h!]
\includegraphics[width=1.0\textwidth]{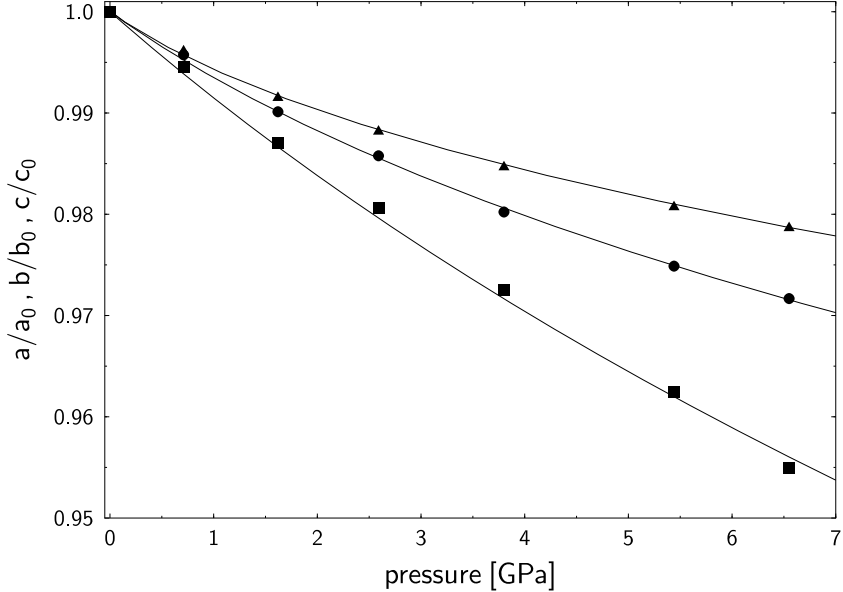}
\caption{\label{lattice-para} Evolution of the normalised lattice
  parameter with pressure. The compression in the three lattice
  directions $a$ ($\blacksquare$), $b$ ($\bullet$) and $c$
  ($\blacktriangle$) is anisotropic. The experimental error
  bars for the pressure and the lattice parameters correspond to the
  size of the symbols. The lines are curve fittings of 3$^{rd}$-order
  Birch-Murnaghan equation-of-states to the data.}
\end{figure}

\begin{figure}[h!]
\includegraphics[width=1.0\textwidth]{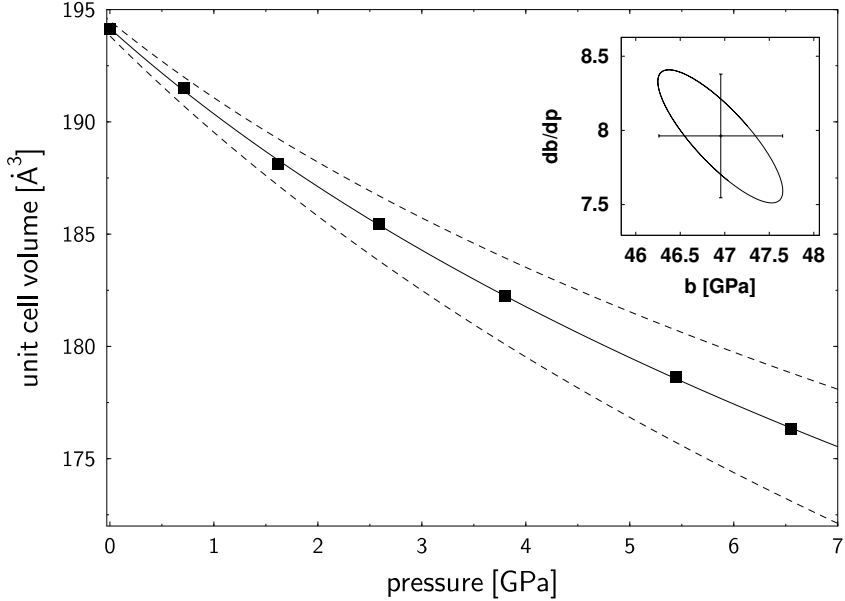}
\caption{\label{eos-fit}Pressure dependence of the unit cell volume of
  $\alpha$-PbF$_2$. The solid line results from the fit of a 3$^{rd}$-order
  Birch-Murnaghan equation-of-state, 
  dashed lines represent the 95~\% confidence region of the fitted
  equation-of-state. The experimental error bars correspond to the size of the
  symbols. The insert shows the one-sigma error ellipsoid of 
  b$_0$ and b' for the fit.}
\end{figure}

As an example the observed and calculated diffraction patterns
resulting from the Rietveld refinement of the 3.8(1)~GPa data are
presented in Figure \ref{powder-fit}. The pressure dependence of the
normalised lattice parameters $a/a_0$, $b/b_0$, $c/c_0$ and the unit
cell volume are shown in Figures \ref{lattice-para} and \ref{eos-fit},
respectively. In Table \ref{hp-lattice} the lattice parameters and the
residuals resulting from the Rietveld refinement at five selected
pressures are given. The refined atomic positions and the resulting
bond distances are given in Tables \ref{hp-structure} and
\ref{bonding}, respectively. The atomic displacement parameters
(U$_{iso}$) were refined to 0.027(3)~\AA$^2$ and 0.029(2)~\AA$^2$ for
lead and the two fluorine ions, respectively

\begin{table}[h]
  \caption{\label{hp-lattice} Residuals, lattice and profile parameters
    from Rietveld refinement for selected pressures.}
  \medskip
  \begin{tabular}{cccccc} \hline
    p [GPa]  & 0.00(4) & 1.62(5) & 3.80(9) & 5.44(9) & 6.55(8) \\ \hline
    R$_{wp}$ \%  & 11.1 & 15.0 & 14.1 & 11.3 & 13.6 \\
    $\chi^2$  & 1.93 & 3.44 & 3.87 & 2.53 & 2.16 \\
    $a$ [\AA] & 6.4567(9) & 6.372(2) & 6.278(1) & 6.213(1) & 6.165(1) \\
    $b$ [\AA] & 3.9071(5) & 3.8680(9) & 3.8294(8) & 3.8088(8) & 3.7963(8) \\
    $c$ [\AA] & 7.666(1) & 7.603(2) & 7.549(2) & 7.520(1) & 7.503(2) \\
    $\eta$ & 0.13(4) & 0.07(3) & 0.25(4) & 0.14(3) & 0.09(4) \\
    U & -0.19(9) & -0.17(5) &  -0.85(2) & -0.53(1) & -0.47(3) \\
    V & 0.14(2) & 0.16(7) & 0.39(3) & 0.29(3) & 0.27(4) \\
    W & 0.0034(3) & 0.006(2) & 0.017(2) & 0.010(2) & 0.009(5) \\ \hline
  \end{tabular}
\end{table}

\begin{table}[h]
 \caption{\label{hp-structure}Atomic positions resulting from 
   Rietveld refinements.}
 \medskip
 \begin{tabular}{lcccccc} \hline
   & \multicolumn{2}{c}{Pb} & \multicolumn{2}{c}{F(1)} & \multicolumn{2}{c}{F(2)} \\
   & {\it x} &  {\it z} & {\it x} & {\it z} & {\it x} & {\it z} \\ \hline
   0.00(4) [GPa] & 0.254(3) &  0.1067(5) & 0.891(7) & 0.051(7) & 0.489(8)
   & 0.850(6) \\  
   1.62(5) [GPa] & 0.253(4) & 0.1111(7) & 0.880(9) & 0.03(2) & 0.49(1) &
   0.850(9) \\
   3.80(9) [GPa] & 0.251(5) & 0.1139(8) & 0.89(1)  & 0.02(1) & 0.50(1) &
   0.854(9) \\
   5.44(9) [GPa] & 0.253(4) & 0.1145(6) & 0.888(9) & 0.041(7) & 0.477(9)
   & 0.853(8) \\
   6.55(8) [GPa] & 0.254(4) & 0.1163(7) & 0.884(9) & 0.037(9) & 0.51(1) &
   0.843(9)  \\ \hline
 \end{tabular}
\end{table}

\begin{table}[h]
  \caption{\label{bonding} Bond distances in the  coordination shell 
    around a lead atom (see fig. \ref{structure}. The values are given in \AA.}
  \medskip
  \begin{tabular}{lccccc} \hline
    p [GPa]  & 0.00(4) & 1.62(5) & 3.80(9) & 5.44(9) & 6.55(8) \\ \hline
    Pb-F(1,9) & 2.47(2) & 2.40(1) & 2.35(2) & 2.39(2) & 2.38(2) \\
    Pb-F(2)   & 2.48(2) & 2.53(2) & 2.51(3) & 2.41(3) & 2.60(3) \\
    Pb-F(3,4) & 2.59(3) & 2.51(3) & 2.49(3) & 2.55(2) & 2.40(2) \\
    Pb-F(5)   & 2.77(2) & 2.78(3) & 2.88(2) & 2.72(1) & 2.72(3) \\
    Pb-F(6,7) & 3.12(3) & 3.10(2) & 3.06(3) & 2.98(2) & 3.03(3) \\
    Pb-F(8)   & 2.38(2) & 2.44(2) & 2.36(2) & 2.36(1) & 2.36(2) \\ \hline
  \end{tabular}
\end{table}

The compression of $\alpha$-PbF$_2$ is extremely anisotropic. The
compression parallel to the $a$ axis is 1.5 times larger than that
parallel to the $b$ axis and twice as large as along the $c$
direction. The volume at zero pressure V$_0$=194.14(4)~\AA$^3$, the
bulk modulus b$_0$=47.0(6)~GPa and its pressure derivative
b'=$\partial$b/$\partial$p=7.9(4) were determined from the
equation-of-state fit.

These results compare to V$_0$= 198.48~\AA$^3$, b$_0$=40~GPa and
b'=4.8 derived from quantum mechanical calculations at the density
functional theory level \cite{Lorenzana1997} and with b$_0$=57.87~GPa
and b'=4.98 derived from Hartree-Fock calculations \cite{Costales2000}
within the accuracy usually obtained in such types of calculations
\cite{Winkler1999,Milman2000}.  The bulk modulus for $\alpha$-PbF$_2$
of 117(4)~GPa reported by Hull and Keen \cite{Hull1998} differs
significantly from the value derived in this work. However, this
difference can be attributed to the limited pressure range and the
linear fitting of the volume data for the determination of the bulk
modulus in the work by Hull and Keen \cite{Hull1998}.  The values for
the b$_0$ and b' obtained in this work are in the same order of
magnitude as found for other halides crystallising in the cotunnite
type structure, e.g. PbCl$_2$ with b$_0$=34(1)~GPa and b'=7.4(6)
\cite{Leger1996} and CaCl$_2$ with a b$_0$=51~GPa \cite{Leger1998}.

The changes of the atomic positional parameters in the investigated
pressure range are small.  The Pb-F distances in the tricapped
trigonal prism range from 2.38(2) to 3.12(3)~\AA \ at ambient
conditions (Tab. \ref{bonding}). The mean of the Pb-F distance in the
coordination polyhedron (2.66(2)~\AA) is close to the sum of the ionic
radii \cite{Shannon1976} for Pb$^{2+}$ (1.35~\AA) in ninefold- and
F$^-$ (1.33~\AA) in fourfold-coordination. Under a pressure of
6.55(8)~GPa the average Pb-F distance is reduced to 2.59(2)~\AA. The
individual Pb-F bonds show a different compression behaviour. The
following notation of the fluorine ions is according to fig.
\ref{structure}.  The bonds in the $ab$-plane decrease by 3.64~\%,
7.34~\% and 0.8~\% for Pb-F(1), Pb-F(3) and Pb-F(8), respectively .
The Pb-F(2) bond is increased by 4.62~\%, Pb-F bonds oriented along
the $c$ direction are less compressible. The Pb-F(5) and the Pb-F(6)
bonds are shortened by 1.8~\% and 2.88~\% at 6.55(8)~GPa.

\begin{table}[h]
  \caption{\label{coordination-tab}  
    Evolution of the tricapped trigonal prism coordination polyhedra in
    $\alpha$-PbF$_2$ with pressure. Tabulated are: The mean fluorine
    distance d$_{mF}$ from the centroid,
    the scattering of the fluorine distances around the mean $\sigma$,
    the distance Pb$_{exc}$ between the lead ion and the centroid of
    the nine fluorine ions, the minimal distance d and the
    mean distance per coordinate d/coord between observed and
    embedded ideal polyhedra. The distortion is
    given in the continuous symmetry measure CSM
    \cite{Pinsky1998} and the quotient r$_q$=r$_s$/r$_e$ of the radii around
    the circumscribed sphere (r$_s$) and caps (r$_e$).}
  \medskip
  \begin{tabular}{lccccccc} \hline
    p [GPa] & d$_{mF}$ [\AA] &  $\sigma$ [\AA]& Pb$_{exc}$ [\AA]&  d  [\AA] &
    d/coord [\AA] & CSM & r$_q$ \\ \hline
    0.00(4) & 2.6574  & 0.557  & 0.3049  & 0.8803  & 0.033  & 1.378 & 1.062  \\
    1.62(5) & 2.6324  & 0.564  & 0.3002  & 0.9350  & 0.035  & 1.492 & 1.024  \\
    3.80(9) & 2.6098  & 0.598  & 0.3023  & 1.1464  & 0.042  & 1.859 & 1.003  \\
    5.44(9) & 2.5857  & 0.477  & 0.2458  & 0.8787  & 0.033  & 1.455 & 1.052  \\
    6.55(8) & 2.5767  & 0.627  & 0.2849  & 0.9667  & 0.036  & 1.607 & 1.006  \\
    \hline
  \end{tabular}
\end{table}
 
The structural changes in a compound under high-pressure can be
described by considering the distortion of the respective coordination
polyhedron. Recently, two different approaches have been developed
\cite{Makovicky1998,Pinsky1998} to quantify the distortion of a
polyhedron by comparison of an ideal to the real one.

\begin{figure}[h!]
  \includegraphics[width=0.5\textwidth]{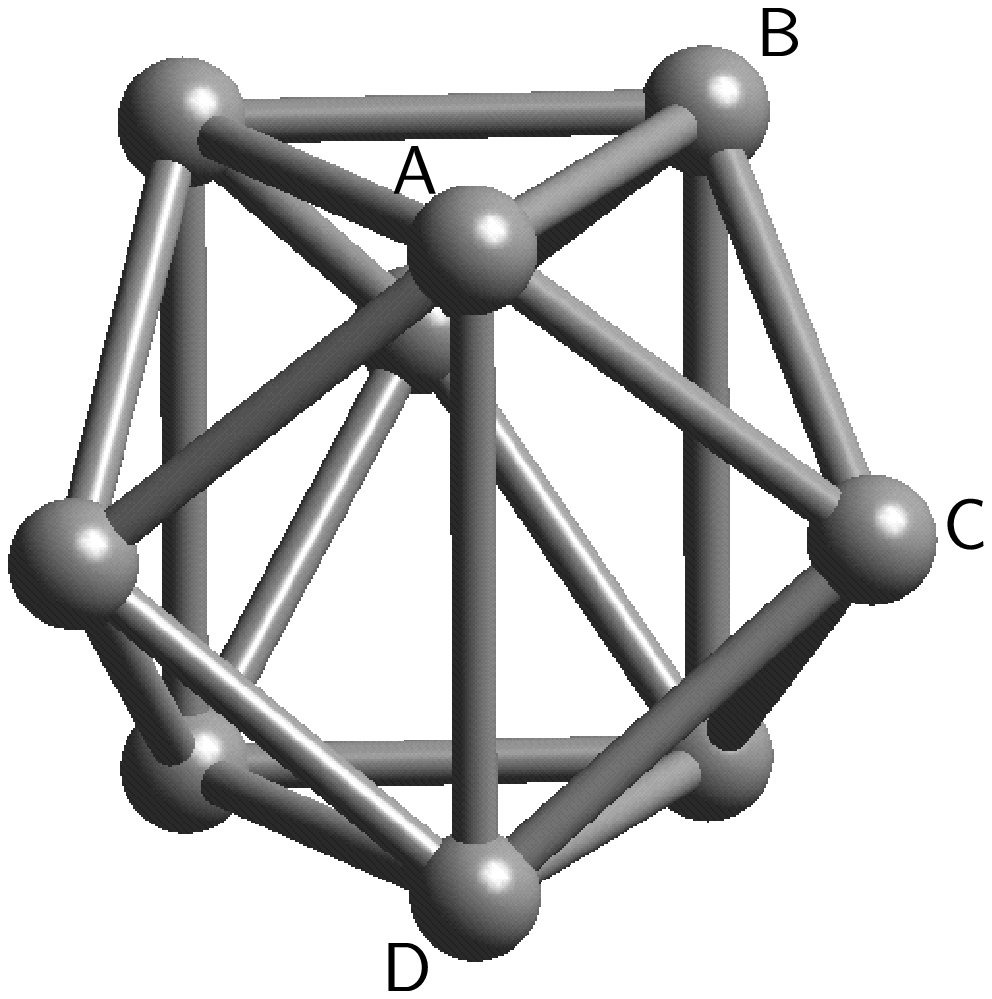}
  \caption{\label{coordination}The regular tricapped trigonal prism with
    all ligands located on a sphere with radius r$_s$. The polyhedral
    edge lengths are AB=AC=2r$_s$/$\sqrt{3}$ and AD=2r$_s$/3$\sqrt5$.}
\end{figure}

The capped square anti-prism and the tricapped trigonal prism (TTP)
are ideal polyhedra for a nine-fold coordination \cite{Kepert1982}.
The TTP is found in $\alpha$-PbF$_2$.  The regular TTP, shown in fig.
\ref{coordination}, is characterised as follows: all nine ligands are
located on the surface of a sphere with radius r$_s$, and the
polyhedra edge lengths are AB=AC=2r$_s$/$\sqrt{3}$ and
AD=2r$_s$/3$\sqrt5$ \cite{Kepert1982}.  The distortion measure of
Mackovicky and Bali\'{c}-\v{Z}uni\'{c} \cite{Makovicky1998} indicates
the volume difference between real and ideal coordination polyhedra.
Therefore the ligands of the real and ideal polyhedra have to be
located on a common circumsphere with radius r$_s$. However, this
measure is not applicable to polyhedra with a large variation in the
bond distances as found in $\alpha$-PbF$_2$.  Hence, the distortion
measure based on the continuous symmetry measure (CSM)
\cite{Pinsky1998} was used.  In this method a regular tricapped
trigonal prism is embedded in the observed one. As an additional free
parameter the radius around the three cap atoms r$_e$ was used to
consider non-spherical deformation.  In order to find the best
conformation, an evolution strategy \cite{Maedler2001} was used to
minimise the euclidean norm (instead of the standard least-squares
method employed by Pinsky and Avnir \cite{Pinsky1998}, which requires
the computation of the two-dimensional derivative with respect to
r$_s$ and r$_e$).  For various pressures the results of the embedding
are given in table~\ref{coordination-tab}.  At ambient pressure the
TTP is slightly distorted (CSM=1.378) compared to the ideal value
(CSM=1). The nine fluorine ions are not arranged on the surface of a
sphere (r$_q$=1) but on an ellipsoid of revolution (r$_q$=1.062).
Remarkably, the lead ion is not located in the conjoined centroid of
the fluorine ligands.  The variance of the fluorine bond distances
increase with pressure, reflecting the difference in the compression
of the individual Pb-F bonds. The lead ion shifts towards the
TTP-centroid, and the arrangement of the fluorine ligands becomes more
spheric with pressure.  However, the minimal distance between the
observed and the embedded fluorine ions increases. Compared to TTP the
distortion is larger and the spheroidal arrangement of the fluorine
ligands extends, both effects can be attributed to increasing angular
distortions.

Raman spectroscopy measurement above 14.7~GPa in PbF$_2$ indicate a
structural phase transition to a post-cotunnite structure with
monoclinic symmetry \cite{Lorenzana1997} as observed in the isotype
SnCl$_2$ and PbCl$_2$ compounds \cite{Leger1996}. The extrapolation of
the CSM to higher pressures indicate a increasing distortion of the
TTP polyhedron. However, from the present CSM data a saturation of the
CSM may not be excluded. Such a behaviour would suggest a ridged-unit
like behaviour of the TTP as it can be found e.g. for the tetrahedral
building units in silicates. Since the structure of the post-cotunnite
phase is not known for PbF$_2$, a further examination of the
high-pressure behaviour seems to be justified.
 
\section{Acknowledgements}
\label{thanks}

LE, KK and HV would like to thank the European Union for financial
support under the TMR program. We are greatful to Korth Kristalle GmbH
Kiel for the donation of the sample and to the German Science
Foundation (DFG) for funding under grant Wi-1232/10.

\end{document}